\documentclass[aps,preprint,showpacs]{revtex4}
\usepackage{graphicx}
\usepackage{dcolumn}
\usepackage{bm}
\usepackage{mathrsfs}

\newcommand{\tauiso}{{\mbox{\boldmath $\tau$}}}

\begin{document}
\title{Hypernuclear production by ($\gamma ,K^+$) reaction within a 
relativistic model} 
\author{R. Shyam$^{1,2}$, H. Lenske$^2$, and U. Mosel$^2$}
\affiliation{$^1$Saha Institute of Nuclear Physics, Kolkata, India\\
$^2$Institute f\"ur Theoretische Physik, Universit\"at Giessen,
D-35292 Giessen, Germany}

\date{\today}

\begin{abstract}
Within a fully covariant model based on an effective Lagrangian picture,  
we investigate the hypernuclear production in photon-nucleus interaction
on $ ^{16}$O target. The explicit kaon production vertex is described via
creation, propagation and decay into relevant channel of $N^*$(1650),  
$N^*$(1710), and $N^*$(1720) intermediate baryonic resonance states in the
initial interaction of the incident photon with one of the target protons.
Bound state nucleon and hyperon wave functions are obtained by solving the
Dirac equation. Using  vertex parameters determined in the previous studies,
contributions of the $N^*$(1710) baryonic resonance dominate 
the total production cross sections which are found to 
peak at photon energies below 1 GeV. The results show that photoproduction
is the most appropriate means for studying the unnatural parity hypernuclear
states, thus accessing the spin dependence of the hyperon-nucleon interaction. 
\end{abstract}
\maketitle

Lambda hypernuclei are the most extensively studied hypernuclear systems,
both experimentally (see, e.g., a recent review~\cite{tam06}) as well
as theoretically~\cite{ban90,nem02,kei00}. They have, traditionally, been 
produced by stopped as well as in-flight ($K^-,\pi^-$) and ($\pi^+,K^+)$
reactions. Alternatively, $\Lambda$-hypernuclei can also be produced with  
proton as well as electromagnetic probes. The feasibility of hypernuclear
production with proton beams has been investigated in Refs.~\cite{kin98,shy04}.
Recently, discrete hypernuclear states have been produced for the first time 
in electron induced reactions on light nuclear targets at the 
Jlab~\cite{miy03,yua06,iod07}. First measurements of the ($\gamma,K^+$) 
reaction on a nuclear target ($^{12}$C) have been reported long 
ago~\cite{mae94,yam95}. Interest in this field has been revived
as a number of experiments are planned for this reaction at accelerators 
MAMI-C in Mainz, ELSA in BONN and also at Jlab (see, e.g., Ref.~\cite{poc05}).
In this paper we report a theoretical study of the ($\gamma,K^+$) reaction on
a $^{16}$O target.

In contrast to the hadronic reactions [($K^-,\pi^-)$, and ($\pi^+,K^+$)]
which are confined mostly to the nuclear surface due to strong absorption 
of both $K^-$ and $\pi^\pm$, the $(\gamma,K^+)$ reaction occurs deep in the 
nuclear interior due to weaker interactions of both photon and $K^+$ with
the nucleus. This property makes this reaction an ideal tool for studying 
the deeply bound hypernuclear states provided the corresponding production 
mechanism is reasonably well understood. Unlike the hadronic reactions which 
excite predominantly the natural parity hypernuclear states, both unnatural and 
natural parity states are excited with comparable strength in the 
$(\gamma,K^+)$ reaction. This is because sizable spin-flip amplitudes are 
present in the elementary $p (\gamma,K^+) \Lambda $ reaction due to the fact
that the photon has spin 1 and small angles dominate this reaction. This 
feature persists in the hypernuclear photoproduction. Furthermore, Since, a
proton in the target nucleus is converted into a hyperon, this reaction leads 
to the production of neutron rich hypernuclei (see, e.g., Ref.~\cite{bres05}) 
which may carry exotic features such as a halo structure.

Several theoretical investigations of the ($\gamma,K^+$) reaction on
nuclear targets have been reported in the literature 
\cite{ber81,shi83,jos88,ben89,mot94,lee95,lee01}. In these studies,
the kaon photoproduction amplitudes on nuclei are calculated within an 
impulse approximation by determining expectation values of the operator 
for the elementary $p (\gamma,K^+) \Lambda$ production process between 
initial and final states of the reaction. This operator is constructed 
either by using the Feynman diagrammatic approach where graphs corresponding
to Born terms and resonance terms in $s$ and $u$ channels, are included
\cite{shi83,jos88,ade85,lee01}, or phenomenologically by parameterizing the 
experimental cross sections for the elementary process~\cite{mot94,lee95}. 
Although in Ref.~\cite{ben89}, Dirac spinors have been used for bound state 
wave functions in the initial and final channels, a full covariant calculation
of this reaction is still missing.
 
In this paper, we study the $A(\gamma,K^+){_{\Lambda}}B$ reaction within a
fully covariant model by retaining the field theoretical structure of the 
interaction vertices and by treating the baryons as Dirac particles moving 
in a static nuclear mean-field. This type of approach has previously been 
used in Ref.~\cite{shy04} to describe the hypernuclear production in 
proton-nucleus collisions. In our model, the initial state interaction 
of the incoming photon with a bound proton leads to excitations of 
$N^*$(1650) [$\frac {1}{2}^-$], $N^*$(1710)[$\frac{1}{2}^+$], and 
$N^*$(1720) [$\frac{3}{2}^+$] baryonic resonance intermediate states which 
are shown to make predominant contributions to $p (\gamma,K^+) \Lambda $ 
cross sections in similar effective Lagrangian studies~\cite{pen02,shk05}. 
Terms involving nucleon intermediate states (Born terms) have not 
been considered here as their contributions are found~\cite{lee01} to be 
insignificant to both elementary as well as in medium (on $^{12}$C nucleus) 
photo-kaon production reactions. 
\begin{figure}
\includegraphics[width=0.15 \textwidth]{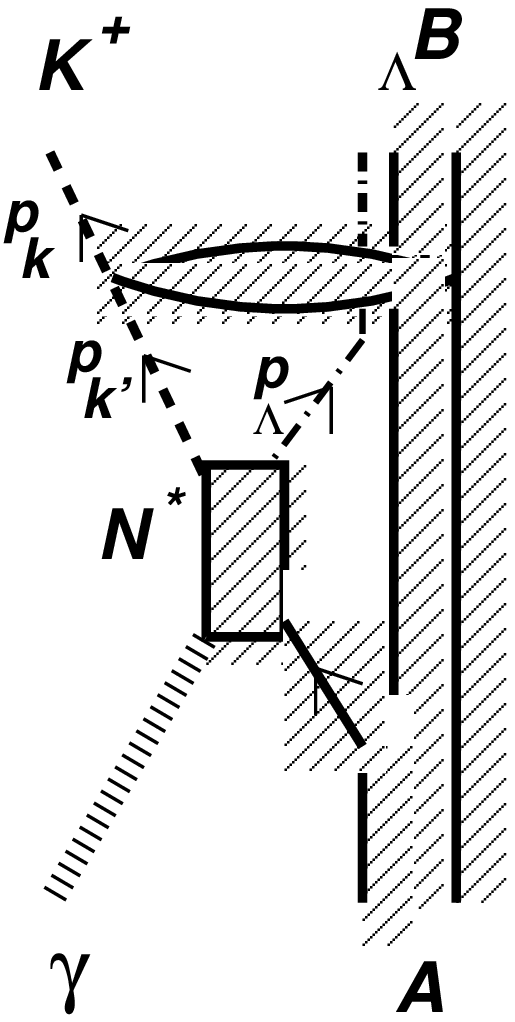}
\caption{Representation of the type of Feynman diagrams included in our
calculations. The elliptic shaded area represents the optical model 
interactions in the outgoing channel.}
\end{figure}
\noindent

We have considered the diagrams of the type shown in Fig.~1 which implies 
that our model has only the $s$-channel resonance contributions. In 
principle, $u$-channel and $t$-channel contributions should also be 
included. However, calculations of the $u$-channel contributions would 
require the knowledge of {\it a priori} unknown couplings to the strange 
baryonic resonances. That is why these graphs have been ignored in recent 
photo-kaon production studies~\cite{pen02,shk05}. The $t$-channel diagrams 
contribute to the non-resonant background and may be requited if the nucleon
intermediate states are considered. In fact, it has been shown 
that~\cite{lee01} for the elementary photoproduction reaction, Born plus 
vector meson $t$-channel contributions become relatively large at photon 
energies in excess of 2.1 GeV~\cite{lee01}. Since, in this paper we have 
concentrated only on investigating the role of baryonic resonances in the 
$(\gamma, K^+)$ reaction and have ignored the nucleon intermediate states, 
we have omitted these terms to keep our model as simple as possible. 
Furthermore, in this exploratory study, to reduce the further computational 
complications we have used plane waves (PW) to describe the relative motion 
of the outgoing kaon which is justified by the relatively weaker mutual 
interaction in this channel.  

The effective Lagrangians for the electromagnetic couplings of   
spin-$1 \over 2$ resonances are given by~\cite{feu98}
\begin{eqnarray}
{\cal L}_{N_{1/2}^*N \gamma} & = & ({eg_{N^*N\gamma}^1 \over {4m_N}})
             {\bar{\Psi}}_{N^*} {\Gamma_{\mu \nu}}
             {\Psi}_N F^{\mu \nu} + h.c.,
\end{eqnarray}
where the operator $\Gamma_{\mu \nu}$ is $\gamma_5 \sigma_{\mu \nu}$
($\sigma_{\mu \nu}$) for odd (even) parity resonances. 
$F_{\mu \nu}$ represents the electromagnetic field tensor: 
$F_{\mu \nu} = \partial_\nu A_\mu - \partial_\mu A_\nu$. We have used 
the notations of Ref.~\cite{bjo64} through out in this paper. For the 
spin-${3 \over 2}$ case, we have
\begin{eqnarray}
{\cal L}_{N_{3/2}^*N \gamma} & = & ({eg_{N^*N\gamma}^1 \over {2m_N}}) 
{\bar{\Psi}}_{N^*}^\alpha \Theta_{\alpha \mu}(z1) \gamma_\nu 
               {\Gamma}{\Psi}_N F^{\mu \nu} \nonumber \\ & - & 
  ({eg_{N^*N\gamma}^2 \over {4m_N^2}}){\bar{\Psi}}_{N^*}^\alpha 
  \Theta_{\alpha \mu}(z2) {\Gamma}({\partial_\nu \Psi}_N) F^{\mu \nu}
\nonumber \\ & + & h.c.,
\end{eqnarray}
In Eq.~(2), the operator $\Gamma$ is unity and $\gamma_5$ for odd and even 
parity resonances, respectively. ${\Psi}_{\mu}^{N^*}$ is the vector spinor
for the spin-${3 \over 2}$ particle. This involves the off shell projector
$\Theta_{\alpha \mu}(z)\,=\, g_{\alpha \mu} - 
{1 \over 2}(1 + 2z)\gamma_\alpha \gamma_\mu$ where $z$ is the off-shell 
parameter~\cite{feu98} which describes the off-shell admixture of 
spin-${1 \over 2}$ fields. The choice of this parameter is arbitrary and
in earlier studies it has been treated as a free parameter to be determined
by fitting to the data (see, e.g.~\cite{pen02}). For a more detailed 
discussion we refer to~\cite{pen02,shy99,pas95}. The electromagnetic coupling
constants $g_1, g_2$ are related to the helicity couplings [$A_{{1/2},{3/2}}]$
(see, e.g. Ref~\cite{feu98}) which are taken from Ref.~\cite{pen02}. There 
these are determined in a coupled channels K-matrix method by fitting 
simultaneously to all the available data for transitions from $\gamma N$ to 
five meson-baryon final states, $\pi N$, $\pi \pi N$, $\eta N$, $K\Lambda$, 
and $K\Sigma$ for center of mass energies ranging from threshold to 2 GeV 
including all the baryonic resonances up to spin $\leq \frac{3}{2}$ and 
excitation energies 2 GeV. 

For the resonance-hyperon-kaon vertices we have  
\begin{eqnarray}
{\cal L}_{N_{1/2}^*\Lambda K^+} & = & -g_{N^*_{1/2}\Lambda K^+}
                          {\bar{\Psi}}_{N^*} {i\Gamma^\prime} \tauiso
                        {\bf \Phi}_{K^{+}} \Psi
                        + {\rm h.c.}.\\
{\cal L}_{N_{3/2}^*\Lambda K^+} & = & \frac{g_{N^*_{3/2}\Lambda K^+}}{m_{K^+}}
                         {\bar{\Psi}}_{\mu}^{N^*} \Gamma
                         {\tauiso} \cdot \partial ^{\mu}
                         {\bf \Phi}_{K^{+}} \Psi + {\rm h.c.}.
\end{eqnarray}
In Eq.~(4), the operator $\Gamma^\prime$ is $\gamma_5$ (unity) for even (odd)
parity resonance.  Signs and values of the hyperon-resonance-kaon coupling 
constants ($g_{N^*YK}$) have been taken from \cite{shy06} which along with 
the helicity amplitudes are shown in Table I. Values of $g_{N^*YK}$ used by
us are also consistent with those reported in Ref.~\cite{pen02}. It may be 
noted that we have used a pseudoscalar (PS) coupling for the 
resonance-hyperon-kaon vertex. Differences between pseudovector and PS 
couplings are expected to be small~\cite{pen02,ben87}. 
\begin{table}
\caption{Coupling constants for the $N^*\Lambda K$ vertices and the helicity 
amplitudes used in the calculations~\protect\cite{shy06,pen02}.
}
\begin{ruledtabular}
\begin{tabular}{cccc}
vertex & $g$ & $A_{1/2}$ & $A_{3/2}$\\ 
                   &   & \footnotesize {($10^{-3} GeV^{-1/2}$)} 
&\footnotesize {($10^{-3} GeV^{-1/2}$)}\\
\hline
$N^*(1710)N\gamma$ & --- & 44 & --- \\
$N^*(1710)\Lambda K^+$ & 6.12 & --- & --- \\
$N^*(1650)N\gamma$ & --- & 49 & ---\\
$N^*(1650)\Lambda K^+$ & 0.76 & --- & --- \\
$N^*(1720)N\gamma$ & --- & 18 & -19 \\
$N^*(1720)\Lambda K^+$ & 0.87 & --- & --- \\
\hline
\end{tabular}
\end{ruledtabular}
\end{table}
The propagators for spin-${ 1 \over 2}$ and spin-${3 \over 2}$ resonances
are taken to be the same as those described in Ref.~\cite{shy04}.

After having established these ingredients, one can  write down, by following
the well known Feynman rules, the amplitudes for graphs of the type shown in
Fig.~1. We have employed pure single-particle-single-hole ($(\Lambda N^{-1})$)
wave functions to describe the nuclear structure part because configuration 
mixing terms are expected to be small. The nuclear structure part is treated
exactly in the same way as is described in Ref.~\cite{ben89}. Amplitudes 
involve momentum space four component (spin space) Dirac spinors ($\psi$) 
which represent wave functions of nucleon and hyperon bound states 
\cite{shy95} and the momentum space kaon-nucleus wave function 
[$\phi_{K}^{(-)*}(p_K^\prime, p_K)$] which can be calculated by using an 
appropriate $K^+$-nucleus optical potential (see, {\it e.g.}, 
Ref.~\cite{tab77}). Momentum $p_K$ represents the asymptotic free state while 
$p_K^\prime$ is the momentum coordinate of the produced kaon as shown in 
Fig.~1. In the PW approximation, one writes 
$\Phi_{K}^{(-)*}(p_K^\prime, p_K)  =  \delta^4(p^\prime_K - p_K)$. 
We include energy dependent widths in denominators of resonance 
propagators to account for the fact that they have finite lifetime for  
decay into various channels.

Spinors $\psi(p)$ are solutions of the Dirac equation in momentum space for
a bound state problem in the presence of an external potential field
\cite{shy95,shy04}
\begin{eqnarray}
p\!\!\!/\psi(p) & = & m_N\psi(p) + F(p),
\end{eqnarray}
where
\begin{eqnarray}
F(p) & = & \delta(p_0 - E) \Biggl[\int d^3p^\prime V_s(-{\bf p}^\prime)
\psi({\bf p} + {\bf p}^\prime) \nonumber \\ & - & \gamma_0
 \int d^3p^\prime V_v^0(-{\bf p}^\prime) \psi({\bf p} + {\bf p}^\prime)
                   \Biggr] .
\end{eqnarray}
In Eq.~(6), the real scalar and timelike vector potentials $V_s$ and $V_v^0$
represent, respectively, the momentum space local Lorentz covariant 
interaction of single nucleon or $\Lambda$ with the remaining $(A-1)$ 
nucleons. We denote a four momentum by $p = (p_0,{\bf p}$). The magnitude 
of the three momentum ${\bf p}$ is represented by $k$, and its directions 
by ${\hat p}$. $p_0$ is the time like component of $p$. Spinors 
$\psi(p)$ and $F(p)$ are written as
\begin{eqnarray}
\psi(p) & = & \delta(p_0-E){{f(k) {\mathscr Y}_{\ell 1/2 j}^{m_j} (\hat p)
                   \choose {-ig(k)}{\mathscr Y}_{\ell^\prime 1/2 j}^{m_j}
                     (\hat p)}}, \nonumber \\
F(p) & = & \delta(p_0-E){{\zeta(k) {\mathscr Y}_{\ell 1/2 j}^{m_j} (\hat p)}
                \choose {-i\zeta^\prime(k){\mathscr Y}_{\ell^\prime 1/2 j}
                     ^{m_j} (\hat p)}},
\end{eqnarray}
where $f(k)$[$\zeta(k)$] is the radial part of the upper component
of the spinor $\psi(p)$[$F(p)$]. Similarly $g(k)$[$\zeta^\prime(k)$] are
the same of their lower component. $f(k)$ and $g(k)$ represent  
Fourier transforms of radial parts of the corresponding coordinate 
space spinors. $\zeta(k)$ are related to $f$, $g$ and the scalar and vector
potentials (see Ref.~\cite{shy95} for more details). We have defined 
$\ell^\prime = 2j - \ell$ with $\ell$ and $j$ being the orbital and total
angular momenta, and 
\begin{eqnarray}
{\mathscr Y}_{\ell 1/2 j}^{m_j}(\hat p) & = & \sum_{m_\ell \mu} 
<\ell m_\ell 1/2 \mu | j m_j> Y_{\ell m_\ell}(\hat p) \chi_{1/2 \mu},
\nonumber 
\end{eqnarray}
where $Y$ represents the spherical harmonics and  $\chi_{1/2 \mu}$ 
the spin space wave function of a spin-$\frac{1}{2}$ particle.

Since our analysis is carried out all along in the momentum space, it 
includes all the nonlocalities in the production amplitude that arise from
the resonance propagators. The differential cross section for the 
$(\gamma,K^+)$ reaction is given by
\begin{eqnarray}
\frac{d\sigma}{d\Omega} & = & \frac{1}{4(2\pi)^2}
\frac{m_Am_B}{(E_\gamma + E_A)^2} \frac{p_K}{p_\gamma}
 \sum_{M_iM_f\epsilon} |\sum_R T_{M_iM_f,\epsilon}|^2, \nonumber 
\end{eqnarray}
where $E_\gamma$ and $E_A$ are the total energies of incident
photon and the target nucleus, respectively while $m_A$ and $m_B$ are
the masses of the target and residual nuclei, respectively. $\sum_R$
represents summation over all the resonances.  $M_i$ and $M_f$ are initial
and final spin states, respectively, and $\epsilon$ is the photon polarization. 

We have chosen the reaction $^{16}$O$(\gamma,K^+)$$^{16}\!\!\!_\Lambda$N for
the first numerical application of our model, as this reaction is well
known to have a simple structure. The initial state is a doubly closed system. 
The binding energies (BE) of $1s_{1/2}$, $1p_{3/2}$ and $1p_{1/2}$ single 
particle states in $^{16}\!\!\!_\Lambda$N hypernucleus are taken to be 
13.20 MeV, 2.50 MeV and 2.04 MeV, respectively which are the same as those
given in Ref.~\cite{ben89}. The BE of the $1p_{3/2}$ and $1p_{1/2}$ nucleon
hole states in $^{16}O$ are taken as 18.40 MeV and 12.13 MeV, 
respectively~\cite{mot94}. Because of the large separation in the binding 
energies of the nucleon hole states, the hypernuclear spectrum is clearly 
divided into 4 groups corresponding to configurations $[p_{1/2}^{-1}, 
s_{\Lambda}]$, $[p_{3/2}^{-1}, s_{\Lambda}]$, $[p_{1/2}^{-1}, p_{\Lambda}]$,
and $[p_{3/2}^{-1}, p_{\Lambda}]$. The configuration mixing is negligible 
except may be for the $J^\pi = 1^+$ hypernuclear state. Any $\Lambda N$ 
residual interaction that may lead to configuration mixing as considered 
in Ref.~\cite{ros88} is neglected in our study.

Spinors $\psi(p)$ for bound hypernuclear and nucleon states  
are obtained by Fourier transformations of the corresponding coordinate
space spinors which are determined by solving the Dirac equation with
scalar and vector potential fields ($V_s$ and $V_v$, respectively) with a 
Woods-Saxon radial form with (reduced radius) $r_s = r_v = 0.983$ fm, and
(diffuseness) $a_s = 0.70$ fm and $a_v = 0.58$ fm. For a given state, the 
depths of these fields ($V_s^0$ and $V_v^0$) have been searched so as to 
reproduce the BE of that state. For the single particle $\Lambda$ states 
$1s_{1/2}$, $1p_{3/2}$ and $1p_{1/2}$ the values of ($V_s^0$ and $V^0$) were
(-204.78 MeV and 165.89 MeV), (-228.49 MeV and 185.96 MeV), and (-251.80 and
203.96 MeV), respectively, while for each of the nucleon hole states $1p_{3/2}$
and $1p_{1/2}$, they were (-445.0 MeV and 360 MeV). We see that the strengths
of the mean field self energies for the $\Lambda$ states are about 1/2 of 
those of the nucleon states which is line with the results of the 
microscopic Dirac-Bruckner calculations as shown in Ref.~\cite{kei00}.  

It is shown in Ref.~\cite{shy95} that spinors calculated in this way provide
a good description of the experimental nucleon momentum distributions for 
various nucleon orbits.  Fig.~2 shows, e.g., the momentum space 
$1p_{1/2}$ $\Lambda$ and $1p_{3/2}$ $\Lambda$ hyperon spinors as functions 
of the momentum transfer ($q$), for the $^{16}\!\!\!_\Lambda$N hypernucleus. 
In the lower panel of Fig. 2 we show the momentum distribution of the 
$\Lambda$ hyperon for these states in $^{16}\!\!\!_\Lambda$N.
We note that only for $q < 1.5 fm^{-1}$, is the magnitude of the lower 
component ($|g(q)|$) substantially smaller than that of the upper component
($|f(q)|$). In the region of $q$ pertinent to the kaon production, 
$|g(q)|$ may not be negligible. In fact, it has been shown 
earlier~\cite{ben89} that the relativistic effects resulting from the small
component of Dirac bound states are large for the kaon photoproduction 
reactions on nuclei.
\begin{figure}
\begin{center}
\includegraphics[width=0.25 \textwidth]{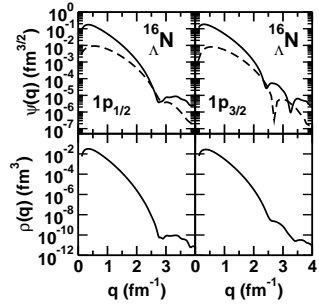}
\caption{ (Upper panel) Magnitudes of the upper($|f(q)|$)(full line) and lower 
($|g(q)|$)(dashed line) components of the momentum space spinors for 
$1p_{3/2}$ $\Lambda$ and $1p_{1/2}$ $\Lambda$ orbits in 
$^{16}\!\!\!_\Lambda$N hypernucleus. (Lower panel) hyperon momentum 
distributions (defined as $\rho(q) = [|f(q)|^2 + |g(q)|^2]$ ) for the same
states.}
\end{center}
\end{figure}

The threshold for the kaon photoproduction on $^{16}$O is about 680 MeV.
The momentum transfer involve in this reaction at zero degrees is above 500
MeV/c. In Fig.~3, we investigate the contributions of the various 
resonance intermediate states to the total cross section for populating the 
$(1p_{3/2}^{-1},1s_{1/2}^\Lambda) 2^-$ state in the 
$^{16}$O$(\gamma,K^+)$$^{16}\!\!\!_\Lambda$N reaction as a function of
photon energy. We see that the individual contributions of the $N^*(1710)$
intermediate state by far dominate the cross sections. In comparison to this,
cross sections corresponding to the $N^*(1650)$ state are at least one order
magnitude smaller and those of the $N^*(1720)$ state are smaller by 3-4 orders
of magnitude (these contributions are omitted from this figure consequently).
The nearly negligible contribution of this resonance has also been noted
in the study of $A(p,K^+){_{\Lambda}}B$ reaction~\cite{shy04}. It is 
worthwhile to note that in Ref.~\cite{lee01} individual contributions 
of various resonances are not specified although the resonance terms as a 
whole are shown to dominate both elementary and nuclear photo-kaon 
production cross sections. 
  
Another noteworthy aspect of Fig.~3 is that cross sections peak at photon 
energies around 900 MeV, which is about 200 MeV above the production 
threshold for this reaction. Interestingly, the total cross section of 
the elementary $p(\gamma,K^+)\Lambda$ reaction also peaks about the same 
energy above the corresponding production threshold (~910 MeV). 
\begin{figure}
\begin{center}
\includegraphics[width=0.25 \textwidth]{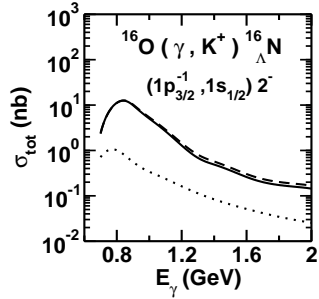}
\caption{ Individual contributions of $N^*$(1710) (dashed line) and $N^*$(1650)
(dotted line) resonance states to the total cross section for 
the $^{16}$O$(p,K^+)$$^{N}\!\!\!_\Lambda$N reaction for the indicated 
configuration as a function of photon energy. The solid line shows the 
coherent sum of the contributions of these two and $N^*(1720)$ resonance
states. The individual contributions of the $N^*(1720)$ resonance
are 3-4 four orders of magnitude smaller and are not shown here separately. }
\end{center}
\end{figure}
Although the absolute magnitude of our near peak energy cross sections 
are comparable to those reported in Ref.~\cite{ben89}, their drop off with
the beam energy is faster than those of these authors. Such a strong drop 
of the cross sections with beam energy beyond the peak position was also 
seen in the calculations of $(p,K^+)$ reactions on nuclei within a 
similar model. It may partly be due to lack of the kaon-nucleus distortion 
effects in our calculations. It should be mentioned here that calculations 
performed with only the upper component of the bound state is about 
6-8$\%$ smaller than the total cross sections shown in Fig.~3. Larger 
effects of the lower components are seen in the angular distributions 
at larger angles.  However, since cross sections are quite small at 
these angles in comparison to those at smaller ones, the total cross 
sections shows less sensitivity to the lower component. More comprehensive 
investigation of the relativistic effects due to the lower component is 
made in Refs.~\cite{ben89} where it is shown that the cross sections 
obtained by using fully relativistic small component are significantly 
different from those where the small component is related to the large 
component by a free space relation. It is also found that the differences 
between cross sections obtained by using Dirac or Schr\"odinger solutions 
for the upper component are non-negligible.
   
In Fig~4, we present the excitation spectrum for four groups of 
$(\Lambda N^{-1})$ hypernuclear states involving bound $1s$ and $1p$ $\Lambda$ 
orbitals in the hypernucleus $^{16}\!\!\!_\Lambda$N. The relative excitation
strengths for a given $J$ state of each group of  
$[(n\ell j)^{-1}_p, (n\ell j)^\Lambda]$ configuration, are obtained by 
dividing the total cross section of that $J$ state by that of the state 
having the maximum cross section within that group. We first note that within
each group the highest $J$ state is most strongly excited which is in  
line with the results presented in Refs.~\cite{ben89,ros88}. Furthermore, 
unnatural parity states within each group [except for the ground state 
$(1p_{1/2}^{-1},1s_{1/2}^\Lambda)$ configuration], are preferentially
excited by this reaction.  For example, within the group 
$(1p_{3/2}^{-1},1p_{3/2}^\Lambda)$, cross section of the $3^+$ state is larger 
than that of the $2^+$ state by about a factor of 2.5 and by more than an 
order of magnitude than that of the $0^+$ state. The unnatural parity states
are excited through the spin flip process. Thus, kaon photoproduction on nuclei
is an ideal tool to investigate the structure of unnatural parity hypernuclear 
states. The addition of unnatural parity states to the spectrum of 
hypernuclei is expected to constrain the spin dependent part of the
effective $\Lambda-N$ interaction more tightly. 
\begin{figure}
\begin{center}
\includegraphics[width=0.30 \textwidth]{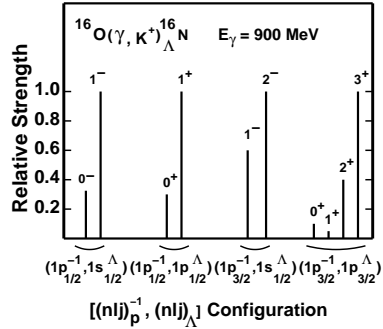}
\caption{ Bound state excitation spectrum of hypernucleus 
$^{16}\!\!\!_\Lambda$N. }
\end{center}
\end{figure}

In summary, we studied the hypernuclear production by $(\gamma,K^+)$ 
reaction on $^{16}$O within a covariant model where in the initial collision 
of the photon with a target proton, $N^*(1710)$, $N^*(1650)$ and $N^*(1720)$
baryonic resonances are excited which subsequently propagate and decay into a
$\Lambda$ hyperon which gets captured in one of the nuclear orbits, and 
a $K^+$ which goes out. Wave functions of nucleon and $\Lambda$ bound 
states are obtained by solving the Dirac equation with appropriate potential
fields. The distortion effects in the $K^+$ channel have not been included in
this study. However, as shown in Refs.~\cite{ben89,ros88}, these effects are 
weak for reactions on $p$-shell nuclei but they may be more significant for 
heavier systems. Since, we have not included the nucleon intermediate states 
(Born terms), the absolute magnitudes of our total cross sections may be 
uncertain to the extent of about 10$\%$.  

Using the vertex constants determined in previous studies, the excitation of
$N^*(1710)$ resonance dominates the hypernuclear production process. Similar
results were also found in the previous studies of the $A(p,K^+){_\Lambda}B$
reaction. The total production cross sections peak at photon energies which
are above the corresponding production threshold by almost the same amount 
of energy as is the position of the maximum in the elementary cross section 
away from its respective threshold.   

Our calculations confirm that the $(\gamma,K^+)$ reaction on nuclei 
selectively excites the high spin unnatural parity states, which makes 
it an ideal tool for investigating the spin-flip transitions 
which are only weakly excited in reactions induced by hadronic probes. 
Therefore, electromagnetic hypernuclear production provides a fuller 
knowledge of hypernuclear  spectra and will impose more severe constraints 
on the models of the $\Lambda N$ interaction, particularly on its the poorly
known spin dependent part. To this end it is important to extend our model 
to include distortion effects in the final channel so that the
mechanism of this reaction can be understood more properly.

This work has been supported by Sonderforschungsbereich/Transregio 16,
Bonn-Giessen-Bochum of the German Research Foundation.
One of the authors (RS) acknowledges the support of Abdus Salam International
Centre for Theoretical Physics in form of a senior associateship award. 

\end{document}